%% file: levelmerge.tex
\documentclass[10pt,journal,compsoc]{./sty/IEEEtran}
\usepackage{cite}
\usepackage[pdftex]{graphicx}

\usepackage{booktabs}

\input{00_macros}

\begin{document}
\title{LevelMerge: Collaborative Game Level Editing by Merging Labeled Graphs}
\author{Christian~Santoni \quad Gabriele~Salvati \quad Valentina~Tibaldo \quad Fabio~Pellacini
\IEEEcompsocitemizethanks{\IEEEcompsocthanksitem Christian Santoni, Gabriele Salvati, Valentina Tibaldo and Fabio Pellacini are in the Computer Graphics research group at the Department of Computer Science, "Sapienza" University of Rome, Italy.}}
%\protect\\
%\protect\\
% note need leading \protect in front of \\ to get a newline within \thanks as
% \\ is fragile and will error, could use \hfil\break instead.
%Website: http://design.di.uniroma1.it/}}

% The paper headers
\markboth{}
{}
\IEEEtitleabstractindextext{%

\input{01_abstract}

% Note that keywords are not normally used for peerreview papers.
}

% make the title area
\maketitle

% To allow for easy dual compilation without having to reenter the
% abstract/keywords data, the \IEEEtitleabstractindextext text will
% not be used in maketitle, but will appear (i.e., to be "transported")
% here as \IEEEdisplaynontitleabstractindextext when compsoc mode
% is not selected <OR> if conference mode is selected - because compsoc
% conference papers position the abstract like regular (non-compsoc)
% papers do!
\IEEEdisplaynontitleabstractindextext
% \IEEEdisplaynontitleabstractindextext has no effect when using
% compsoc under a non-conference mode.

% For peer review papers, you can put extra information on the cover
% page as needed:
% \ifCLASSOPTIONpeerreview
% \begin{center} \bfseries EDICS Category: 3-BBND \end{center}
% \fi
%
% For peerreview papers, this IEEEtran command inserts a page break and
% creates the second title. It will be ignored for other modes.
\IEEEpeerreviewmaketitle

\ifCLASSOPTIONcompsoc
\IEEEraisesectionheading{\section{Introduction}\label{sec:introduction}}
\fi
% Computer Society journal (but not conference!) papers do something unusual
% with the very first section heading (almost always called "Introduction").
% They place it ABOVE the main text! IEEEtran.cls does not automatically do
% this for you, but you can achieve this effect with the provided
% \IEEEraisesectionheading{} command. Note the need to keep any \label that
% is to refer to the section immediately after \section in the above as
% \IEEEraisesectionheading puts \section within a raised box.

% The very first letter is a 2 line initial drop letter followed
% by the rest of the first word in caps (small caps for compsoc).
% 
% form to use if the first word consists of a single letter:
% \IEEEPARstart{A}{demo} file is ....
% 
% form to use if you need the single drop letter followed by
% normal text (unknown if ever used by IEEE):
% \IEEEPARstart{A}{}demo file is ....
% 
% Some journals put the first two words in caps:
% \IEEEPARstart{T}{his demo} file is ....
% 
% Here we have the typical use of a "T" for an initial drop letter
% and "HIS" in caps to complete the first word.

\IEEEPARstart{G}{ame} level editing is the process of constructing a full game level starting from 3D asset libraries, e.g. 3d models, textures, shaders, scripts. In level editing, designers define the look and behavior of the whole level by placing objects, assigning materials and lighting parameters, setting animations and physics properties and customizing the objects AI and behavior by editing scripts. The heterogeneity of the task usually translates to a workflow where a team of people, experts on separate aspects, cooperate to edit the game level, often working on the same objects (e.g.: a programmer working on the AI of a character, while an artist works on its 3D model or its materials). Today this collaboration is established by using version control systems designed for text documents, such as Git \cite{progit}, to manage different versions and share them amongst users. The merge algorithms used in these systems though does not perform well in our case since it does not respect the relations between game objects necessary to maintain the semantic of the game level behavior and look. This is a known problem and commercial systems for game level merging exists, e.g. PlasticSCM \cite{plasticscm}, but these are only slightly more robust than text-based ones. This causes designers to often merge scenes manually, essentially reapplying others edits in the game level editor.

\begin{figure*}[ht!]
  \centering
\includegraphics[width=\linewidth]{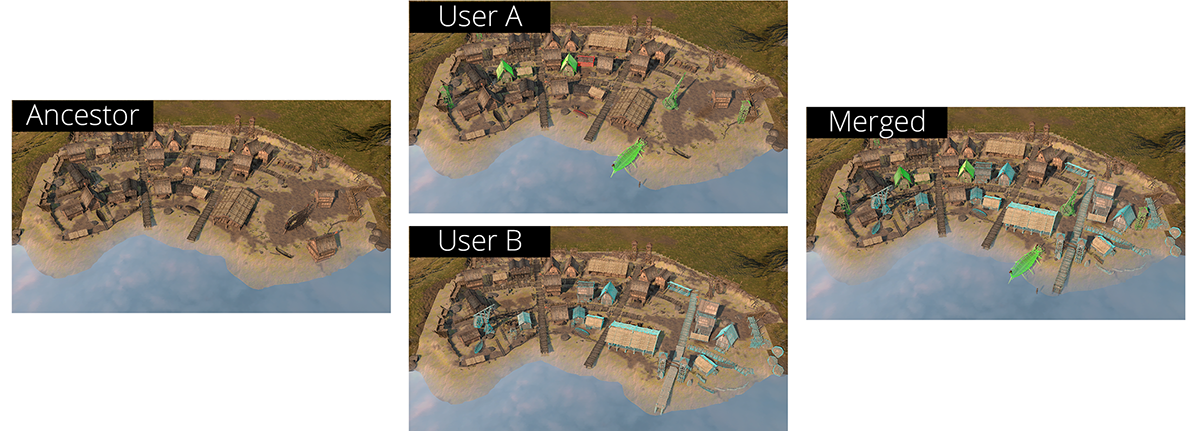}
  \caption{Example merge computed with our algorithm. The original scene (\emph{left}) is concurrently edited by two artists (\emph{middle}), and automatically merged by our algorithm (\emph{right}). Our merge algorithm maintains the semantic of user edits making the game playable after merging. The edits are color coded: green for User A, cyan for User B, red for dropped edits}
  \label{fig:teaser}
\end{figure*}

\input{02_intro}

\input{03_related}
\input{04_diff_merge}

\input{05_collab_system}
\input{06_results}

\input{07_conclusions}

\bibliographystyle{IEEEtran}
\bibliography{./levelmerge}
\end{document}

%% file: 00_macros.tex
\newcommand{\parasection} [1] {\textsf{\textbf{#1}.}}

\newcommand{\oldversion } [1] { }
\newcommand{\ignorethis } [1] { }

\newcommand{\tblnum     } [1] {\ref{#1}}
\newcommand{\fignum     } [1] {\ref{#1}}

\newcommand{\tbl        } [1] {Table~\tblnum{#1}}

\newcommand{\fig        } [1] {Fig.~\fignum{#1}}
\newcommand{\Reals      }     {{\textrm{I\kern-0.18em R}}}

\newcommand{\change     } [1] {\mbox{{\footnotesize $\Delta$} \kern-3pt}#1}

%% file: 01_abstract.tex
% !TEX root =  levelmerge.tex
Game development is commonly seen as a collaborative effort, with teams of people cooperating on the same project. Nowadays, a variety of cloud-based services have shown the benefits of performing tasks in real-time collaboration with others. In this paper, we present a system for collaborative game levels editing. We model this problem as a special instance of merging labeled directed acyclic graphs. We propose an algorithm that guarantees that the shared game level is always coherent between edits, both hierarchically and semantically.
We establish real-time collaboration by initiating merges automatically and by augmenting the game editor interface to allow users to monitor all others' edits in real-time. We validate our algorithm by merging complex edit and large game levels. We further validate the collaborative workflow by running a user study with expert game developers. It shows that our system works well and collaborative workflows are beneficial to game development.

%% file: 02_intro.tex
% !TEX root =  levelmerge.tex

\parasection{Collaborative level editing}
The goal of this paper is to present a system for collaborative game level editing in the style of cloud-based services such as Google Docs or Clara.io \cite{claraio}. Just like in these services, we want designers to edit the level concurrently, while viewing others' edits as they are applied. From a system perspective, we differ from these commercial efforts in two main manners. First, we build our collaborative editor on an existing game engine, in our case Unity \cite{unity}, rather than using a simplified, instrumented interface. This allows us to create complex and playable games concurrently. Second, to ensure compatibility with current editing workflows, we maintain the use of distributed version control, in our case GIT, as a low-level component to manage versions and share them amongst users over a network.

\parasection{LevelMerge}
To support collaborative workflows, we provide a robust merge algorithm for game levels, we establish collaboration by automatically merging edits from multiple users, and we augment the game editor interface to show the component that all users are editing. This results in a system for collaborative game level editing, with which users work on the same level concurrently and seamlessly. Said another way, the game is always playable during the concurrent editing session. We show this in the supplemental video with sped-up collaborative editing sessions on playable game levels.

Our system works by diffing and merging a shared game level. During merges, we maintain a coherent state between concurrent edits, both \emph{hierarchically}, by preserving the hierarchical relations in the game level, and \emph{semantically}, by ensuring that concurrent edits do not break data dependencies between entities in the scene, e.g. scripts that refer to assets or animations that refer to rigged models. In the rare instances of conflicts in parts of the scene, i.e. when specific edits to game objects cannot be merged automatically, designers can either resolve the conflict manually within the editor, or use an automatic conflict resolution algorithm that applies edits from a preferred branch, chosen by the user. These simple policies proved to be intuitive to understand and effective to use in our testing. In our system conflicts are rare for two reasons. First, since our diff algorithm is very precise, we have no false positives. Second, we augment the interface to show where other artists are working. This in turn mean that artists naturally avoid most conflicts while working. This is similar to Google Docs showing selections and cursors for all users.

\parasection{Results}
We tested our prototype implementation by merging large playable game levels after significant edits created in Unity \cite{unity}. One such merge is shown in \fig{fig:teaser}. 
We tested game levels between $205$ and $3172$ MB with complex object hierarchies of between $79$ and $2800$ nodes. The merge algorithm ran in between $0.07$ and $0.92$ seconds, fast enough to support collaboratively workflows. We allow users to change the sync interval interactively since sync may require large scene changes that might be undesirable while editing.

We validated our system running an user study with $6$ game developers, already Unity users, of which $3$ where experts designers with game publishing experience. The study showed that (1) our system was preferred over the traditional version control workflows, (2) even in small scenes, our merge algorithm was significantly better than test-based merging with manual conflict resolution, and (3) all users felt that this manner of collaboration is a significant improvement in creating games.

\parasection{Contributions}
To summaries, in this paper we present a system for collaborative game level editing. In our opinion, our work has four main contributions. (1) We demonstrate collaborative game level editing with a workflow that is compatible with real game engines. (2) We propose a merge algorithm for game levels that maintains the hierarchical and semantic relationships between game objects during merges, leading to merges where the game stays consistently playable. (3) We propose modification to game editors interfaces to allow for collaboration. (4) We show with a user study that collaborative workflows are useful in game editing.

%% file: 03_related.tex
% !TEX root =  levelmerge.tex

\section{Related work}

The use of version control is a well-established practice in a variety of fields, especially for text based documents. For 3D assets, these methods are becoming only recently available. For 3D meshes, both \cite{3ddiff} and \cite{meshgit} make use of a version control approach, developing systems able to diff, merge and resolve conflicts between different versions of 3D models. \cite{nonlinear_image} shows a version control formulation for image editing. We utilize these methods to merge single assets in our scenes. The works that are mostly related to our system we are about to present are the online services Clara.io \cite{claraio} and OnShape \cite{onshape}. Although presenting a system for real-time collaborative workflows, they only focus on 3D polygonal modeling and CAD modeling scenes, respectively. We instead aim to provide real-time syncing at the whole game level granularity.

\parasection{Game level version control}
Scenes version control framework is presented in \cite{3Drevcontr}. In this approach, the whole scene is represented by a graph and every 3D asset is treated as an atomic blob of binary data. Concurrent edits on the same scene node are detected as a conflict even if the edits are not overlapping at asset level. Moreover, this approach doesn't take into account the semantic of the objects in the scene, an aspect that is necessary to guarantee coherence between merges of a game level. As for game scenes' versioning, to the best of our knowledge, while integration with version control systems exists, only one commercial product, PlasticSCM \cite{plasticscm}, attempt to add semantic during merging. While the merge algorithm is not explicitly documented, our tests suggested that it doesn't take into consideration the semantic relationship between objects, essentially performing a classical line-by-line diff instead. We instead aim to present a system that considers the whole game level as a labeled directed acyclic graphs, performing all diff, merge and conflict resolution operations on such data structure, maintaining the hierarchical and semantic coherence of the scene.

\parasection{Labeled graphs}
For their intrinsic hierarchical structure, graphs are commonly used to represent game levels. The literature shows in fact multiple examples of works that adopted this approach. \cite{xml3Drepo} presents an optimized API for game scenes' versioning, encoding all the objects and assets into an unified scene graph. Though, since its implementation is based on the versioning architecture of \cite{3Drevcontr}, this framework can't manage the semantic aspects of merging a game level. \cite{serviceOriented} presents an architecture capable to expose a web service interface to the users, to review and control 3D scenes. The core of the system is based on openSG \cite{opensg}, an open source C++ scene graph. In this case the graph structure is used mostly for performance-related issues, and the framework doesn't offer any features useful to versioning or sharing game levels.

%% file: 04_diff_merge.tex
% !TEX root =  levelmerge.tex

\begin{figure}[ht!]
	\centering
    	\includegraphics[width=0.47\textwidth]{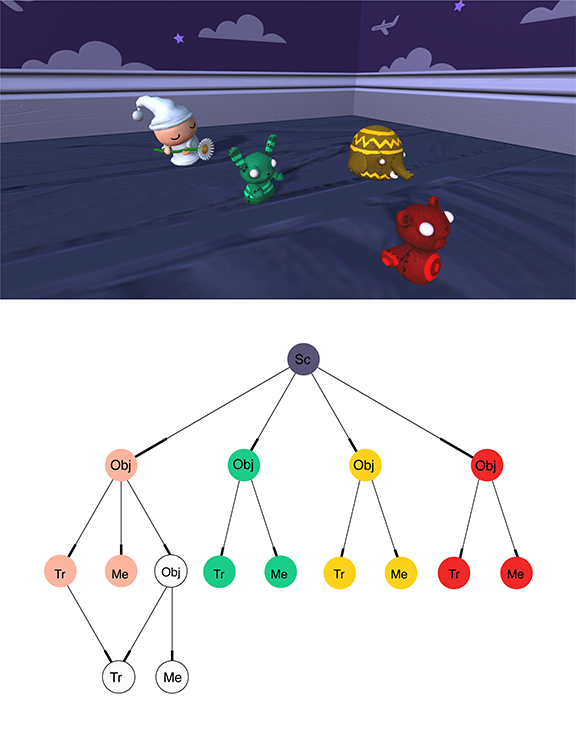}
	\vspace{-0.1in}
    	\caption{Example game level and corresponding labeled directed graph. We only show nodes corresponding to the foreground characters. Nodes in the graph are colored accordingly to the color of the object in the level, and are labeled with the abbreviation of the node's type.}
        \label{fig:scene_example}
\end{figure}

\section{Diff and merge algorithm}
\label{diff_merge}

\parasection{Game Level}
In our system, a game level is composed of a set of objects, such as animated characters, lights, and cameras. Each object has associated components whose properties specify the object's look and behavior such as transform and material nodes, collision proxies and code such as shaders and AI scripts. Objects datas are specified by referencing assets that are not edited directly in the level, such as rigged meshes, textures, animation data, etc. Said another way, a game level is an heterogenous collection of entities of different types that are related to each other.

\parasection{Level Model}
To define a merge algorithm we need an abstract model of a game level that works across the heterogeneity of level entities. In our algorithm, we model game levels as labeled direct acyclic graphs (LDAGs), where nodes are game objects and assets. Each node is specified by a unique identifier, and a set of single-valued properties (e.g transforms position, asset parameters). We have two types of edges, expressing whether or not there is a direct dependency between parent and child. For direct dependencies, changes in the parents' nodes are mirrored to the children, while for indirect dependencies changes need not be mirrored.  \fig{fig:scene_example} shows a simple scene with its associated graph. We show edges with both direct dependencies (e.g.: a container object hierarchically including all the various components of a 3D mesh), and indirect dependencies (e.g.: a script with all the assets it instantiates, a particle system with its generated objects). This definition allows to capture the heterogeneity of object and their relationships in game levels and it is where we differ most from standard scene graphs that are either purely hierarchical or purely data-driven.

\parasection{Asset Merging}
In our system we treat assets separately since they cannot be edited directly in the game engine interface; this implies that we do not explicitly model 3D meshes, textures or code in the graph. Rather, our system can be seamlessly integrated with algorithms that are specifically designed to merge each kind of assets, e.g. \cite{meshgit} for 3D models, \cite{progit} for code, \cite{nonlinear_image} for images. One exception we make is with code assets that are not merged if they do not compile since this would break the playability of the game level. This was found useful also in collaborative code editing.

\begin{figure*}[htb!]
	\centering
    	\includegraphics[width=\textwidth]{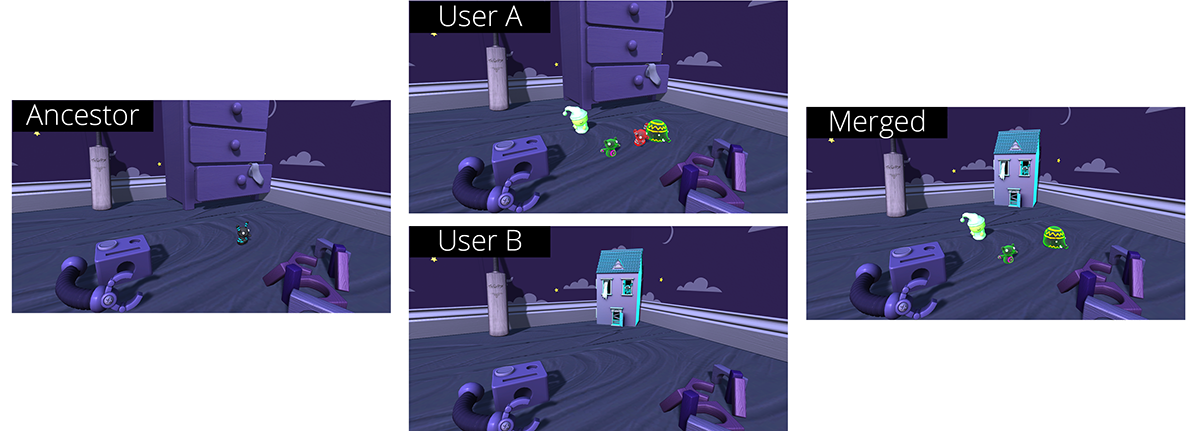}
    	\caption{Example merge with dependency changes. Both users add new objects to the scene and User B changes bunny's direct dependency, making it child of the doll house. User B also deleted the drawers: since it is not in direct dependency to any modified object, is deleted.} 
        \label{fig:merge}
\end{figure*}

\parasection{Level Merging}
The merging procedure follows the classic 3-way diff approach, and is thus based on the analysis of three separate graphs: the user's current scene graph, the shared upon remote scene's graph and the common ancestor graph. We compute the difference between an edited version and its ancestor, stored explicitly in our system, by marking nodes as either unchanged or added, deleted or modified with respect to the ancestor graph. This can be done efficiently since nodes have unique identifiers in our system. A node is considered to be modified if any of its properties have been modified or if any of its parents in direct dependency has been modified. 

We thus obtain the two difference graphs of edited nodes (by the user and in respect to the remote scene) as the subgraphs induced by the edited nodes on the levels graph. We compute the merged result by applying the modifications as a whole to the ancestor graph, while detecting if any modification is in conflict. We proceed by considering the cases of added, deleted and modified nodes respectively. While doing so, we maintain the level consistent by ensuring that the merged graph is still an LDAG. An example of a level merge without conflicts is shown in \fig{fig:merge}.

Nodes that are simply added in one of the difference graphs can be inserted automatically in the ancestor graph (since adding nodes cannot create conflicts). In this case, a new edge is added to the resulting graph between the node and the parent node in direct dependency with it. If the node has no parents in the scene, the added edge is between the node and the scene-root node.

For node deletions, a preliminary check must be performed, in the case such node was also modified by another user. If this doesn't happen, the node can be removed safely, and its subtrees are linked to the deleted node's parent, to avoid generating disconnected components. Instead, if the node is modified in the remote graph, then a conflict occurs. Moreover, deletions affects all nodes in direct dependency with the deleted node, since they would be deleted too. Since those child nodes can be in indirect dependency with other nodes in their subtrees, the deletion operation can induce disconnected components in the graph. Thus, a conflict occurs even in the case that there is a modification for any of the nodes in the subtree induced by the deleted node. The approach we apply when a conflict is detected will be explained in the next section. %In this case we add an edge between the first node in each subtree not in direct dependency and the deleted node's parent, again to avoid creating disconnected components.

We will now discuss merge strategies for modified nodes. Parameter changes are merged comparing single-value in modified nodes, in pairs of the same type. If the same parameters are modified in both nodes, a conflict occurs. Otherwise, both sets of modifications are merged into the node components.
 
Though, applying edits in this manner might induce graphs that contain cycles. After the merge and conflict resolution phases, we thus check their presence. If a cycle is detected, we remove it by sorting the nodes contained in the cycle in terms of height in the LDAG (topological order). We then remove the first edge with lowest height, representing an indirect dependency. In this way we avoid propagation of changes. If there is no indirect edge, we drop the direct one with smallest height.

\begin{figure*}[htb!]
	\centering
    	\includegraphics[width=\textwidth]{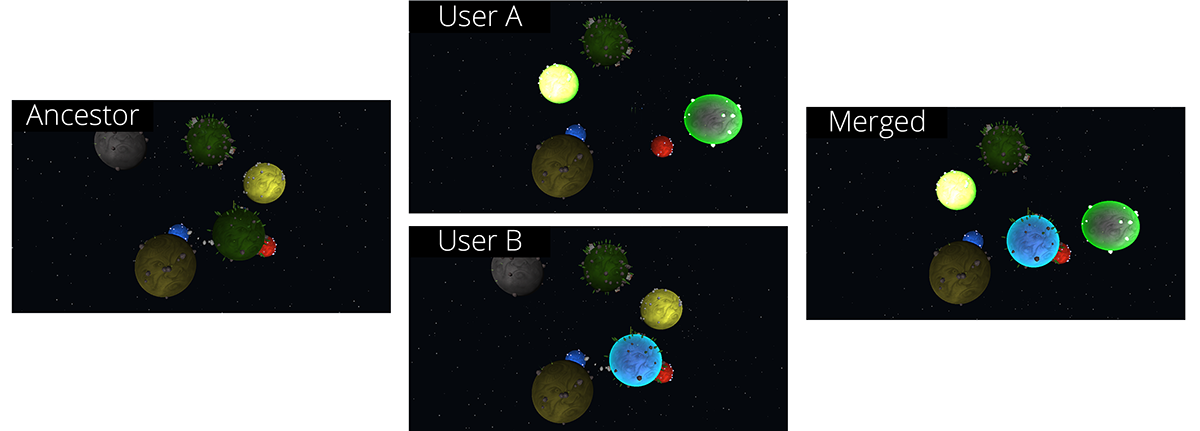}
    	\caption{Example merge in the presence of conflicts. A conflict arises since User B modifies the material associated to the front green planet, while User A deletes it from the scene. In case of conflicts, users can either resolve them manually or set a preference of which branch to maintain. In this case, we resolved the conflict by maintaining branch A. The results show that, accordingly to the chosen policy, the object isn't deleted in the merged scene, and retains the edits done by User A.}
        \label{fig:conflict}
\end{figure*}

\parasection{Conflict resolution}
In summary, conflicts may arise when multiple users modify the properties of the same node or when operations are applied by one user on a node deleted by another. While in these cases the user can resolve the conflict manually, just like in version control, we also provide an automatic conflict resolution algorithm that can apply the edits of a branch over the ones from the other branch, where the user selects the prioritized branch. This policy is common in text-based version control systems, and aims to automatize what users usually do manually, manipulating directly the conflicting files and selectively applying the local edits rather than the remote ones, or vice-versa. An example of how a conflict is automatically resolved (respectively selecting merge policy 1) and 2) ) for a merged scene is shown in Figure \ref{fig:conflict}. In our testing we found that users preferred to auto-resolve conflicts, which were very rare, by accepting edits from the shared repository thus aiding in faster collaboration.

%% file: 05_collab_system.tex
\section{System Implementation and Limitations}

Our implementation leverages on two main systems, Git \cite{progit} as a version control backend and Unity \cite{unity} as a game editor and engine. Here we describe implementation details necessary for reproducing our work. 

\parasection{Integration with Git}
We follow common practice in game version control to handle our collaboration backend. We establish collaboration by having users share a centralized Git repository and use Git to store level versions and as a protocol to efficiently communicate scene diffs. We chose Git since it is a well-know, scalable and robust versioning protocol, and since it allows users to have the whole working tree locally stored, thus allowing them to work locally at full speed, while relying on network transfer only for user-to-user synching. Given the size of graphics assets, which may take several seconds just to download on a local LAN, this last feature is of paramount importance. In Git, users can register specialized diff and merge algorithms that are triggered on different file types. We integrate our merge algorithm by registering it for merging full level files and handle assets by integrating specific algorithms discussed before. For code assets, we selectively synch them to the main repository when they successfully compile,  while maintaining a separate repository for sharing intermediate versions.

\parasection{Integration with Unity}
We implemented our algorithm on Unity3D 5 \cite{unity}, a commercial game engine. We chose a commercial engine since it allows us to test fully-functioning games at scale and since it is widely adopted in the community. Our algorithm though does not depend on Unity per se, since we work on arbitrary labeled graph. Here we describe the steps for integrating our algorithm in the Unity editor. We take a lightweight approach for integration and rather than modifying directly the engine data structures, we act on the serialized levels. For each merge, we parse Unity's serialized format into our own representation, merge the level LDAGs, and rewrite a new scene. We integrate this merge procedure directly in the game editor, by issuing synching actions within the editor itself with a plugin and issuing scene reloads after merges, which happens asynchronously to game editing to avoid locking the interface. Still, Unity does lock during scene reloading, since it has to refresh assets' caches. For this reasons, users felt that it was comfortable to sync only every few seconds to avoid locking the interface during asset reloading for a smoother workflow (note that only happens when strictly necessary).

\begin{figure}[ht!]
	\centering
    	\includegraphics[width=0.45\textwidth]{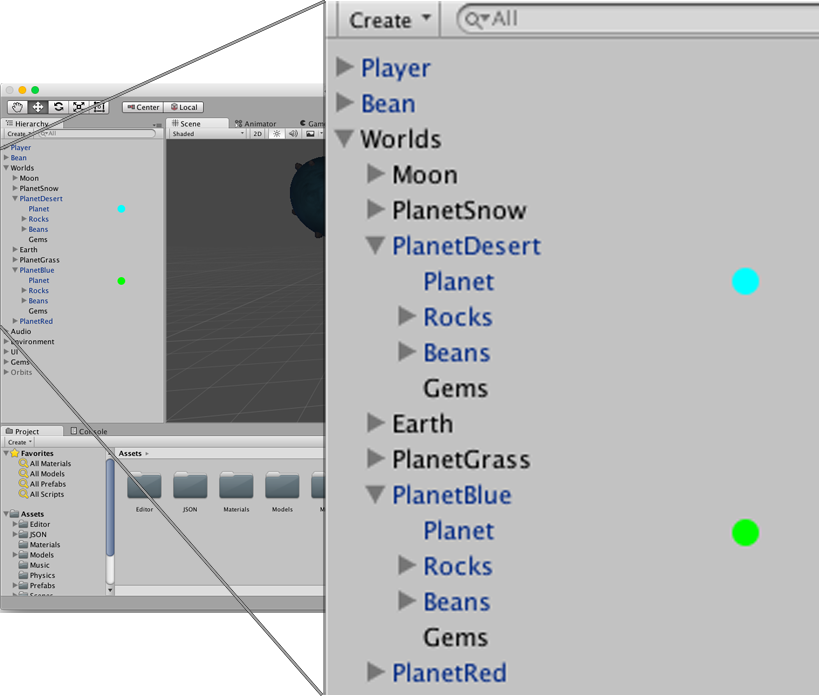}
    	\caption{A screenshot of the extended Unity's interface we present to the users. When the sync system is active, users are notified by colored dots of other users' edits on objects of the scene (on the left, in the hierarchy inspector) and assets (bottom center, in the asset explorer). Every user is uniquely identified by a color.}
        \label{fig:interface}
\end{figure}

As stated before, we also extend the Unity interface to support collaboration between users. As shown in \fig{fig:interface}, we show colored indicators in the ``hierarchy inspector'', for edits done on objects instantiated in the game level, and in the ``asset browser'', for modifications done on the assets, where the color uniquely identify users. We considered applying this approach also in 3D scene viewer, but felt this would not work since coloring the actual objects the scene would hinder materials appearance. In our opinion this feature helps users significantly in understanding the actual development state of the scene and also reduces unwanted conflicts, preventing multiple people to work on the same entities at the same time. 

\parasection{Limitation: Selective conflict resolution}
When editing sequences become long, a user might want the aid of a specialized interface to selectively resolve conflicts, akin to Git ``cherry picking''. Currently, users can resolve conflicts by editing the partially merged scene, or automatically resolve conflicts with our simple policies. In this scenario, a specialized interface might be fruitful.

\parasection{Limitation: Expand assets' merging strategies}
In our system we treat assets as atomic entities, relying on external algorithms for their merges. One asset that we cannot handle currently is merging keyframed animation data since no published algorithm exists for this kind of asset. Furthermore, it may be possible to introduce more mathematically sound manners of merging material properties, instead of averaging their values, by taking advance of linearized editing spaces \cite{appim}.

\parasection{Limitation: UI improvements}
As a result of the informal user study we run, some of the users complained about the fact that our indicators were not sufficiently visible. So, a possible improvement could be to place such indicators directly in the scene, near the currently modified components. Also, it would be useful to notify the user also on the specific components' attributes on which other people are working on.

\parasection{Limitation: On-the-fly modifications in simulation mode}
Finally, our system blocks the syncing algorithm when an user enters "simulation mode", i.e when the game is fully built and the level is played. An interesting extension could be to allow users to all enter the simulation mode and share modifications to the scene while actually playing it. 

%% file: 06_results.tex
% !TEX root =  levelmerge.tex

\begin{table*}[tbh]
\centering
\begin{tabular}{cccccccccc}
\toprule
     & \textbf{ancestor} & \textbf{merge} & \textbf{ancestor} & \textbf{ancestor} & \textbf{diff A} & \textbf{diff B} & \textbf{merged} & \textbf{merged} \\
\textbf{name} & \textbf{size (MB)} & \textbf{time (s)} & \textbf{nodes} & \textbf{edges} & \textbf{nodes} & \textbf{nodes} & \textbf{nodes} & \textbf{edges} \\
\midrule
\emph{room} & 205 & 0.07 & 79 & 84 & 168 & 85 & 244 & 248 \\
\emph{planets} & 149 & 0.80 & 2702 & 3352 & 545 & 31 & 2184 & 2710 \\
\emph{lab} & 3172 & 1.40 & 2800 & 3156 & 131 & 303 & 3142 & 3498  \\
\emph{vikings} & 1620 & 6.16 & 2249 & 2318 & 361 & 467 & 2384 & 2452  \\
\bottomrule
& &  &  &  &  &  &  & 
\end{tabular}
\vspace{0.1in}
\caption{ Summary statistics of merged scenes. Editing time for \emph{vikings} was significantly higher than the other levels, amounting to roughly 1.5 hours per branch.}
\label{tbl:results}
\end{table*}

\section{Results}

We tested our merge algorithm by merging large edits on complex scenes. We implemented our system as an unoptimized Python script running on a 2.6 GHz desktop machine. We purposely tested merging after a significant amount of edits were performed to validate the robustness of our approach. Within normal collaboration, users can choose the frequency of merging. In our testing, we choose to sync every roughly 10 seconds, which works well since in level editing many edits are not necessary as frequent (as they could be, for example, in 3D modelling), especially when using large assets and editing code. The supplemental video shows collaborative editing sessions on one of our scenes, as well as walkthroughs of the final merged versions for the larger scenes (since editing sessions lasted too long).

\subsection{Merge Algorithm}

\begin{figure*}[htb!]
	\centering
    	\includegraphics[width=\textwidth]{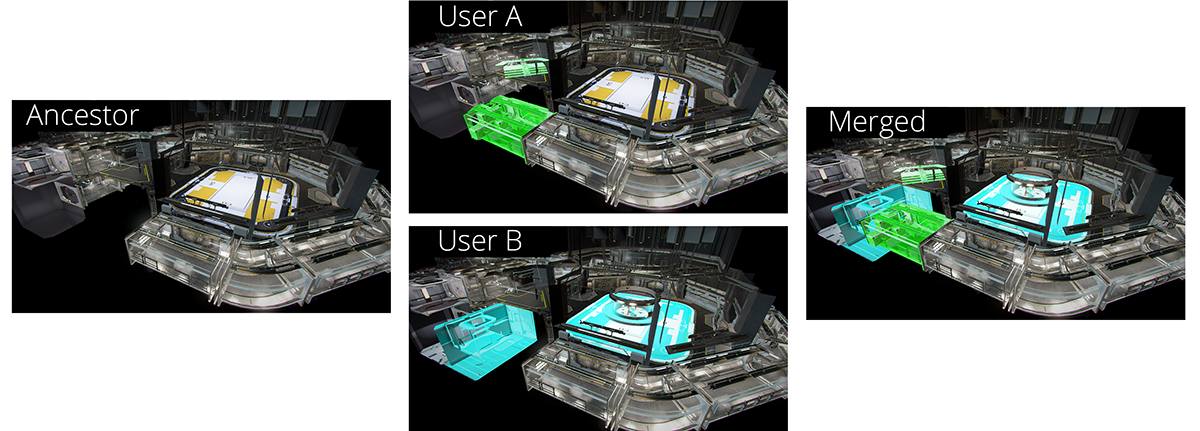}
    	\caption{In this final image, we perform edits to a scene (lab) with a complex graph and large assets to show that our systems scales well with scene complexity. All the information regarding the merge are displayed in \tbl{tbl:results}.}
        \label{fig:final}
\end{figure*}

\tbl{tbl:results} shows statistics on merging large edits in four different game levels, shown in Figs.~\ref{fig:teaser}, \ref{fig:merge}, \ref{fig:conflict} and \ref{fig:final}. Games were playable before and after the merge and no user intervention was involved in the merging. We chose scenes of different complexity in terms of asset size and graph dimensions, from a simple interior scene (\emph{room}) scenes, to a space game with detailed planets (\emph{planets}), from a large first person environment (\emph{vikings}) to a detailed laboratory interior (\emph{lab}). For these scenes, the whole level (including assets) ranged from $149$ to $3172$ MB of data, while the corresponding graphs range from 79 to 2800 nodes.

We performed a variety of edits to the scenes \emph{room}, \emph{planets} and \emph{lab} including adding and deleting objects, changing materials and textures, adding and deleting lights, changing AI code and shaders, modifying physics behavior and navigation meshes.  This corresponds to graph diffs spanning between 31 and 545 nodes, a considerable edit with respect to the original scene. The merge of those scenes took, on average, between 0.07 and 1.4 seconds. The speed of the merge algorithms support collaborative workflow very well, being faster than modifying code or loading large assets. We merged without any manual intervention resolving conflicts by automatically choosing a preferred branch. After the merge, the level maintains all its hierarchical and semantic relationship, and in fact is playable as is.

As a final test, we performed significant modifications to the \emph{vikings} scene, amounting to roughly 1.5 hours per branch. The goal of this test was to stress test the system under the most demanding conditions, rather than in the condition normally used for collaborative workflows. Even in this case our merge algorithm produce a consistent scene, but with a considerable slowdown, taking roughly 6.16 seconds. Note that manually merging this scene in any reasonable time would have been impossible.

\subsection{User Study}

We run a user study with experts and novice Unity users, to measure the goodness of the merge policy and validate the benefits of a more tight collaborative workflow in game development. During the study we gave users both guided and free tasks, to perform in small groups, working in cooperation. In each task we compare two merge workflows: our algorithm vs a text-based workflow (using GIT merge with manual conflict resolution). During the experiment, we provided users with a working game level similar to \emph{room}. We tested a wide range of tasks typical for game editing: scripting, light editing, shaders editing, assets' parameters tuning, texture and materials editing, game logic and game AI scripting.

\parasection{Goal}
The aim of the user study is to measure the goodness of the merge algorithm in terms of user fruition, by testing whether our merge algorithm integrates with the user's editing behaviour in a way that can be approachable for users.

\parasection{Subjects}
We recruited $9$ users, $7$ experts and $2$ novices. We define expert subjects as game designers that have either won a game jam competition, were author of a publication about game design, or have published a game in an app store. The novices were 3D artists with at least 4 years of experience in a company.  In all tasks we ask users to cooperate with other users for completing a common goal, in groups of $2-3$ users each time, without mixing experts and novices.

\parasection{Tasks}
We run four guided task and an open task of about half an hour. Each task had two repetitions, each for a different merge workflow. After each change of merge policy, we ask user to rate the difficulty of completing the task on a scale from $1$ to $5$, where $1$ is ``hard to be completed'', and $5$ is ``very easy to complete''. At the end of each task, we asked users an open feedback about the task.

In the guided tasks, we asked each user to complete actions that we designed, as they can lead the scene to inconsistent states such as cycles, disconnected components or edits drop. We tested the following conflicts resolution strategies: same objects addition, hierarchical conflict, semantic conflict, deletion conflict.  In the open task we asked user to freely edit the game scene, with the must of adding a script. We give them some suggestions about the script, and about the scene resource.

\parasection{Experimental Procedure}
We provide users with a starting playable game level and a variety of assets and ask them to perform various editing tasks. The experiment lasted one and a half hours plus the additional times for training and final survey completion. No talking was allowed during task completion neither before starting tasks. To avoid learning effects, we randomized the order in which the merge algorithms were presented in the various tasks. To avoid ``good subject'' effects, users did not know what was changing between repetitions. Since using Git merge can lead to level graphs that are inconsistent, we allow users to edit directly the scene file or to use Git tools for reverting files in case of conflicts. We left to users the choice to skip a task or to consider it blocking in a way such that the scene can not be recovered. 

\parasection{Results}
We extend the rating scale of the experiment we gave to users, considering with values $0$ the case in which users marked the task as blocking or if they revert the scene via Git, rolling to a previous scene state thus losing all edits. The average rate in term of ``how hard has been to complete task'' for the two samples are $4.11$ for our merge and $1.85$ for git merge. During the experiment, git's merge corrupted the $23.6\%$ of scenes. All users except one of the novices felt the difference between the two merging methods and both novices and experts reported the scene corruption as the worst event happened during task completion. 

\parasection{Informal Feedback}
We collect informal feedback on the project overall and the main concept expressed by users is that this can be an amazing tool for increasing communication and to shorten development time. Here we cite some of the sentences taken from the experts: \textit{"I think this should be the ideal path for making actively communicate different professions increasing quality of results $\dots$ obviously this system would extremely make easier teamwork, decrease deployment time and would be gratifying for teams that will notice realtime system progression"}.

%% file: 07_conclusions.tex
% !TEX root =  levelmerge.tex

\section{Conclusions}

In this paper we presented a system for real-time collaborative game level editing. We solved the problem of merging the levels modeling them as a labeled directed cyclical graphs, and performing all diffing, merging and conflict resolution operations on that data structure. In this way, we guarantee that the merged scene is alway hierarchically and semantically coherent between edits. We present to the user three different policies to automatically merge and solve conflicts. For assets, we provide specific merge strategies for 3D models, eulerian transformations applied to objects, and materials, and we structured our system to be extensible to include new strategies for other kinds of assets. For code sharing, our system periodically performs compile checks, to guarantee that the shared code (and, consequently, the whole project) is always runnable. Finally, we present to the users an interface that notifies them about other people's current edits. We found our system to be reliable, robust and scalable both to scene's complexity and to long edit sessions.

%% file: levelmerge.bbl
% Generated by IEEEtran.bst, version: 1.12 (2007/01/11)
\begin{thebibliography}{10}
\providecommand{\url}[1]{#1}
\csname url@samestyle\endcsname
\providecommand{\newblock}{\relax}
\providecommand{\bibinfo}[2]{#2}
\providecommand{\BIBentrySTDinterwordspacing}{\spaceskip=0pt\relax}
\providecommand{\BIBentryALTinterwordstretchfactor}{4}
\providecommand{\BIBentryALTinterwordspacing}{\spaceskip=\fontdimen2\font plus
\BIBentryALTinterwordstretchfactor\fontdimen3\font minus
  \fontdimen4\font\relax}
\providecommand{\BIBforeignlanguage}[2]{{%
\expandafter\ifx\csname l@#1\endcsname\relax
\typeout{** WARNING: IEEEtran.bst: No hyphenation pattern has been}%
\typeout{** loaded for the language `#1'. Using the pattern for}%
\typeout{** the default language instead.}%
\else
\language=\csname l@#1\endcsname
\fi
#2}}
\providecommand{\BIBdecl}{\relax}
\BIBdecl

\bibitem{progit}
S.~Chacon, \emph{Pro Git}.\hskip 1em plus 0.5em minus 0.4em\relax APress, 2009.

\bibitem{plasticscm}
PlasticSCM, ``Plasticscm documentation,'' \url{http://www.plasticscm.com},
  2015.

\bibitem{claraio}
Clara.io, ``Clara.io documentation,'' \url{http://clara.io}, 2015.

\bibitem{unity}
Unity3D, ``Unity3d documentation,'' \url{https://unity3d.com/}, 2015.

\bibitem{3ddiff}
\BIBentryALTinterwordspacing
J.~Dobo\v{s} and A.~Steed, ``3d diff: An interactive approach to mesh
  differencing and conflict resolution,'' in \emph{SIGGRAPH Asia Technical
  Briefs}, 2012, pp. 20:1--20:4. [Online]. Available:
  \url{http://doi.acm.org/10.1145/2407746.2407766}
\BIBentrySTDinterwordspacing

\bibitem{meshgit}
\BIBentryALTinterwordspacing
J.~D. Denning and F.~Pellacini, ``Meshgit: Diffing and merging meshes for
  polygonal modeling,'' \emph{ACM Trans. Graph.}, vol.~32, no.~4, pp.
  35:1--35:10, 2013. [Online]. Available:
  \url{http://doi.acm.org/10.1145/2461912.2461942}
\BIBentrySTDinterwordspacing

\bibitem{nonlinear_image}
\BIBentryALTinterwordspacing
H.-T. Chen, L.-Y. Wei, and C.-F. Chang, ``Nonlinear revision control for
  images,'' \emph{ACM Trans. Graph.}, vol.~30, no.~4, pp. 105:1--105:10, 2011.
  [Online]. Available: \url{http://doi.acm.org/10.1145/2010324.1965000}
\BIBentrySTDinterwordspacing

\bibitem{onshape}
OnShape, ``Onshape documentation,'' \url{http://www.onshape.com}, 2015.

\bibitem{3Drevcontr}
\BIBentryALTinterwordspacing
J.~Dobo\v{s} and A.~Steed, ``3d revision control framework,'' in \emph{Proc. of
  17th International Conference on 3D Web Technology}, 2012, pp. 121--129.
  [Online]. Available: \url{http://doi.acm.org/10.1145/2338714.2338736}
\BIBentrySTDinterwordspacing

\bibitem{xml3Drepo}
\BIBentryALTinterwordspacing
J.~Dobo\v{s}, K.~Sons, D.~Rubinstein, P.~Slusallek, and A.~Steed, ``Xml3drepo:
  A rest api for version controlled 3d assets on the web,'' in \emph{Proc. of
  International Conference on 3D Web Technology}, 2013, pp. 47--55. [Online].
  Available: \url{http://doi.acm.org/10.1145/2466533.2466537}
\BIBentrySTDinterwordspacing

\bibitem{serviceOriented}
\BIBentryALTinterwordspacing
A.~Schiefer, R.~Berndt, T.~Ullrich, V.~Settgast, and D.~W. Fellner,
  ``Service-oriented scene graph manipulation,'' in \emph{Proc. of
  International Conference on Web 3D Technology}, 2010, pp. 55--62. [Online].
  Available: \url{http://doi.acm.org/10.1145/1836049.1836057}
\BIBentrySTDinterwordspacing

\bibitem{opensg}
D.~Reiners, G.~Voﬂ, and J.~Behr, ``Opensg: Basic concepts,'' in \emph{OpenSG
  Symposium}, 2002.

\bibitem{appim}
F.~Di~Renzo, C.~Calabrese, and F.~Pellacini, ``Appim: Linear spaces for
  image-based appearance editing,'' \emph{ACM Trans. Graph.}, vol.~32, no.~6,
  pp. 194:1--194:9, 2014.

\end{thebibliography}
